\documentclass[twocolumn,english]{revtex4}
\usepackage[utf8]{inputenc}
\usepackage{color}
\usepackage{amsmath}
\usepackage{graphicx}

\makeatletter
\@ifundefined{textcolor}{}
{%
 \definecolor{BLACK}{gray}{0}
 \definecolor{WHITE}{gray}{1}
 \definecolor{RED}{rgb}{1,0,0}
 \definecolor{GREEN}{rgb}{0,1,0}
 \definecolor{BLUE}{rgb}{0,0,1}
 \definecolor{CYAN}{cmyk}{1,0,0,0}
 \definecolor{MAGENTA}{cmyk}{0,1,0,0}
 \definecolor{YELLOW}{cmyk}{0,0,1,0}
}

\makeatother

\usepackage{babel}
\begin{document}

\title{Configuration Dynamics of a Flexible Polymer Chain in a Bath of Chiral
Active Particles}

\author{Xinshuang Liu}
\author{Huijun Jiang}
 \email{hjjiang3@ustc.edu.cn}
\author{Zhonghuai Hou}
 \email{hzhlj@ustc.edu.cn}

\affiliation{Department of Chemical Physics \& Hefei National Laboratory for Physical
Sciences at Microscales, University of Science and Technology of China,
Hefei, Anhui 230026, China}

\date{\today}
\begin{abstract}
We investigate configuration dynamics of a flexible polymer chain in a bath of active particles with dynamic chirality, i.e., particles rotate with a deterministic angular velocity $\omega$ besides self-propulsion,by Langevin dynamics simulations in two dimensional space. Particular attentions are paid to how the gyration radius $R_{g}$ changes with the propulsion velocity $v_{0}$,angular velocity $\omega$ and chain length. We find that in a chiral bath with a typical nonzero $\omega$, the chain first collapses into a small compact cluster and swells again with increasing $v_{0}$, in quite contrast to the case for a normal achiral bath $(\omega=0)$ wherein a flexible chain swells with increasing $v_{0}$. More interestingly, the polymer can even form a closed ring if the chain length is large enough,which may oscillate with the cluster if $v_{0}$ is large. Consequently, the gyration radius $R_{g}$ shows nontrivial non-monotonic dependences on $v_{0}$, i.e., it undergoes a minimum for relatively short chains, and two minima with a maximum in between for longer chains. Our analysis shows that such interesting phenomena are mainly due to the competition between two roles played by the chiral active bath: while the persistence motion due to particle activity tends to stretch the chain, the circular motion of the particle may lead to an effective osmotic pressure that tends to collapse the chain. In addition, the size of the circular motion $R_{0}=v_{0}/\omega$ shows an important role in that the compact clusters and closed-rings are both observed at nearly the same values of $R_{0}$ for different $\omega$. 
\end{abstract}
\maketitle

\section{INTRODUCTION }

As the volume fraction of large macromolecules and other inclusions
in cell interior can be as large as 50\%\cite{fulton1982crowded},
crowded environment has been reported to be a key factor in many life
processes including chromosome organization\cite{pelletier2012physical},
polymer translocation through nanopore\cite{gopinathan2007polymer,palyulin2014polymer},
gene expression\cite{tan2013molecular}, and so on. \textcolor{black}{Specifically,
}configuration change of polymers in crowded environments has gained
extensive research attentions in last decades\cite{foster2009metastable,haydukivska2014ring,hu2000collapse,kim2015polymer,majumder2017kinetics,shin2015kinetics,bian2018understanding,toan2006depletion,shin2015polymer,jeon2016effects,stiehl2013kinetics,eilers1988protein,kim2011crowding,chen2019comparative,wang2017understanding,kang2015effects,prakash2004protein,rodriguez2013multistep,balchin2016vivo,li2009effects},
due to its close relevance to many crucial biological processes\cite{prakash2004protein,rodriguez2013multistep,balchin2016vivo,li2009effects,kang2015effects}.
For instance, compared to the non-crowded environment, \textcolor{black}{a}\textcolor{blue}{{}
}polymer chain has a tendency to collapse into a compact structure
and increase looping probability in a crowded environment\cite{chen2019comparative,jeon2016effects,wang2017understanding,prakash2004protein,eilers1988protein,stiehl2013kinetics,kim2011crowding,kang2015effects,shin2015polymer}.
\textcolor{black}{Such a collapse was found to be dependent on the
size of the crowder, since smaller crowders exert a higher osmotic
pressure onto the polymer and make the polymer chain collapse into
a more compact structure as compared to the bigger ones \cite{kang2015effects}.
In addition, polymer rigidity also played an important role on the
looping characteristics of a linear polymer chain in crowded environment\cite{shin2015polymer},to
list just a few.}

It is noted that, crowders in many similar studies are all assumed
to be passive, i.e., the crowders themselves will finally relax to
an equilibrium state determined by temperature and their interaction
potentials. Nevertheless, the crowders may also be active ones which
spontaneously take up energy from the environment and convert into
their directed motions. Actually, explosive new non-equilibrium phenomena
in active systems have been reported in recent years \cite{czirok2000collective,woolley2003motility,sokolov2007concentration,vicsek2012collective,fily2012athermal,schwarz2012phase,wittkowski2014scalar,stenhammar2015activity,cohen2014emergent,dunkel2013fluid,jiang2017emergence,pu2017reentrant,ding2017study,feng2017mode,feng2018mode,du2019study,bechinger2016active},
both experimentally and theoretically. In particular, dynamics of
polymer in active crowded environments also shows many nontrivial
phenomena\cite{bechinger2016active,kaiser2014unusual,harder2014activity,shin2015facilitation,pu2016polymer,nikola2016active,xia2019globule}.
Kaiser and Löwen found that the polymer extension follows a two-dimensional
Flory scaling for very long chains, while the active crowders can
stiffen and expand the chain in a nonuniversal way for short chains
\cite{kaiser2014unusual}. Harder et al. revealed that, the polymer
collapses \textcolor{black}{into a metastable hairpin for a moderate
activity of the crowders, and reexpands eventually as the activity
increases to be large enough for a rigid polymer chain\cite{harder2014activity}.
Shin et al. investigated the influence of active crowders on polymer
looping, and found that increasing the activity yields a higher effective
temperature and thus facilitates the looping kinetics of the polymer
chain\cite{shin2015facilitation}. }

Besides of the activity, the motion of crowders in real active systems
may also be dynamically chiral \cite{corkidi2008tracking,Riedel2005,lauga2006swimming,kummel2013circular}.\textcolor{blue}{{}
}\textcolor{black}{Examples of chiral active particles including sperm
cells \cite{corkidi2008tracking,Riedel2005}, E.coli swimming in circular
trajectories near a surface \cite{lauga2006swimming}, asymmetric
L-shaped microswimmers of circular motion on a substrate\cite{kummel2013circular},
etc.}\textcolor{blue}{{} }Compared to the active crowder without dynamic
chirality, the chiral one is affected not only by the active force
but also by an extra torque, leading to a circular motion in two dimensional
space or a helical motion in three dimensional space. It has been
found that the chiral feature of active particles can affect significantly
the collective dynamics including transport\cite{liao2018transport,wu2016transport,ai2016ratchet,hu2016transport},
separation\cite{ai2015chirality,liebchen2017collective}, sorting\cite{mijalkov2013sorting,chen2015sorting},
and so on. However, to the best of our knowledge, how the \textcolor{black}{dynamic
chirality of active crowders }would influence the configuration change
of \textcolor{black}{a }polymer chain has not been studied yet.

Motivated by this, in this paper, we investigate the \textcolor{black}{configuration
dynamics o}f a linear polymer chain immersed in an active bath with
dynamic chirality in two dimensional space. Each active particle is
self-propelled with a speed $v_{0}$ along a direction which changes
via random rotational diffusion and besides, a deterministic rotation
with angular velocity $\omega$. We mainly focus on how the averaged
radius $R_{g}$ of gyration of the polymer chain depends on the active
velocity $v_{0}$, with varying $\omega$ and chain length $N$. Very
interestingly, we find that $R_{g}$ shows nontrivial non-monotonic
dependences on $v_{0}$: it first decreases to a minimum value and
then increase again. At the minimum, the chain collapses into a small
compact cluster that may rotate in the opposite direction to that
of the particles. If the chain is long enough, more interesting behaviors,
including a closed ring with some particles rotating inside, and an
oscillation between the cluster and the ring, can be observed for
moderate levels of particle activity. 

The remainder of this paper is organized as follows. In Sec. II, we
describe our model and simulation method. In Sec. III, we present
our results and discussion followed by conclusions in Sec. IV. 

\section{MODEL AND METHOD}

\textcolor{black}{We consider a} linear polymer chain consisting of
$N$ coarse-grained spring beads with diameter $\sigma$ immersed
in a bath of $N_{c}$ chiral active \textcolor{black}{crowder particles}
in a $L\times L$ two dimensional space with periodic boundary conditions.
The crowders are of the same diameters as the beads. All the non-bonded
(excluded volume) interactions among beads and crowders are described
by \textcolor{black}{the} WCA potential,
\begin{equation}
U_{WCA}\left(r_{ij}\right)=\begin{cases}
4\varepsilon\left[\left(\dfrac{\sigma}{r_{ij}}\right)^{12}-\left(\dfrac{\sigma}{r_{ij}}\right)^{6}+\frac{1}{4}\right] & r_{ij}\leq2^{1/6}\sigma\\
0 & r_{ij}>2^{1/6}\sigma
\end{cases},
\end{equation}
where $\varepsilon$ denotes the potential strength, and $r_{ij}$
is the distance between a pair of particles $i$ and $j$, with $i,j$
running over all the crowders and beads. Besides, the bonded interaction
between all neighboring beads in the polymer chain is described by
the finite extension nonlinear elastic (FENE) potential,
\begin{equation}
U_{FENE}(r_{ij})=-\frac{1}{2}\kappa R_{0}^{2}ln\left[\left(1-(\frac{r_{ij}}{R_{0}})^{2}\right)\right]
\end{equation}
where $\kappa$ is the elastic coupling constant, $R_{0}$ denotes
the maximum bond length and $i,j$ run over the polymer beads.

For the active chiral crowders, the system dynamics obeys the following
overdamped Langevin equations

\begin{equation}
d\vec{r_{i}}/dt=v_{0}\overrightarrow{u_{i}}-(1/\gamma)\nabla_{i}U+\sqrt{2D_{0}}\xi_{i}^{T}(t),
\end{equation}
\begin{equation}
d\theta_{i}/dt=\omega+\sqrt{2D_{R}}\xi_{i}^{R}(t),
\end{equation}
where $\vec{r_{i}}$ and $\theta_{i}$ are the position and orientation
of the $i$th crowder, respectively. $\vec{u_{i}}\equiv(cos\theta_{i},sin\theta_{i})$
denotes the orientation of the $i$th active crowder and $v_{0}$
is the active velocity. In particular, $\omega$ in (2) gives the
angular velocity of the particle which is assumed to be the same for
all crowders. For $\omega\ne0$, the particle undergoes a circular
motion in the absence of the noise term, which determines the chirality
of the particle. The model reduces to the usually used ABP (stands
for Active Brownian Particles) one for $\omega=0$. 

In Eq. (1), $U$ denotes the total interaction potential in the whole
system including all the exclusive-volume and bonded interactions,
$\gamma$ is the friction coefficient of the surrounding medium, $D_{0}$
and $D_{R}$ are the translational and rotational diffusion coefficients,
respectively, which hold the relations of $\gamma=k_{B}T/D_{0}$ and
$D_{R}=3D_{0}/\sigma^{2}$ with $k_{B}$ the Boltzmann constant and
$T$ the temperature. $\xi_{i}^{T}(t)$ and $\xi_{i}^{R}(t)$ are
Gaussian white noise\textcolor{black}{s} with zero mean and satisfy
$\left\langle \xi_{i\alpha}^{T}(t)\xi_{j\beta}^{T}(s)\right\rangle =\delta_{ij}\delta_{\alpha\beta}\delta(t-s)$
($\alpha,\beta$ denote Cartesian coordinates) and $\left\langle \xi_{i}^{R}(t)\xi_{j}^{R}(s)\right\rangle =\delta_{ij}\delta(t-s)$,
respectively. 

Note that for a polymer bead $i$, the dynamics is also described
by Eq.(1) except that $v_{0}=0$. For simplicity, we use same interaction
and diffusion parameters for the polymer beads and the crowder particles. 

In simulations, $\sigma$ , $k_{B}T$, and $\tau=\sigma^{2}/10D_{0}$
are chosen as the dimensionless units for length, energy and time,
respectively. Other parameters are chosen as $L=50$, $D_{0}=0.1$,
$D_{R}=0.3$, $\gamma=10.0$, $\varepsilon=10$, $\kappa=300$, $R_{0}=1.5$,
and the simulation time step is $10^{-4}$. The number of crowders
is set to be $N_{c}=318$, so that its volume fraction is given by
$\phi=0.1$. All the data in this paper are averaged over $50$ independent
simulation runs. 

\section{RESULTS AND DISCUSSION }

The main purpose of the present work is to investigate how the chiral
feature of the crowders would influence the configuration dynamics
of the polymer chain. To begin, we consider a chain with fixed length
in a chiral active bath with a fixed nonzero angular velocity, but
varying the activity velocity $v_{0}$. Fig.1 shows several typical
snapshots from our simulations for $N=40$ and $\omega=1.5\pi$, with
increasing $v_{0}$ from a relatively small value $v_{0}=1$ to a
large one $v_{0}=50$. When the activity is small as shown in Fig.1(a),
the polymer chain keeps a free state (the free state here means that
the chain is neither stretched nor collapsed), similar to a polymer
chain in a passive environment. With increasing activity to a moderate
level as shown for $v_{0}=15$ in Fig.1(b), interestingly, the polymer
chain collapses into a stable compact spiral structure. All the crowder
particles move outside this cluster and collide with the cluster.
Consequently, the whole cluster rotates in the opposite direction
(clockwise) to that of the crowder particles. For further increasing
$v_{0}$, for instance $v_{0}=30$ in (c), crowders may intrude into
the interior of the cluster leading to its breakup and the chain becomes
more expanded than the cluster. If the activity is even larger, the
crowders would aggregate and adhere to the chain leading to stretching
of the chain, as depicted for $v_{0}=50$ in Fig.1(d). 

\begin{figure}
\centering{}\label{fig.snapshot40}\includegraphics[scale=0.7]{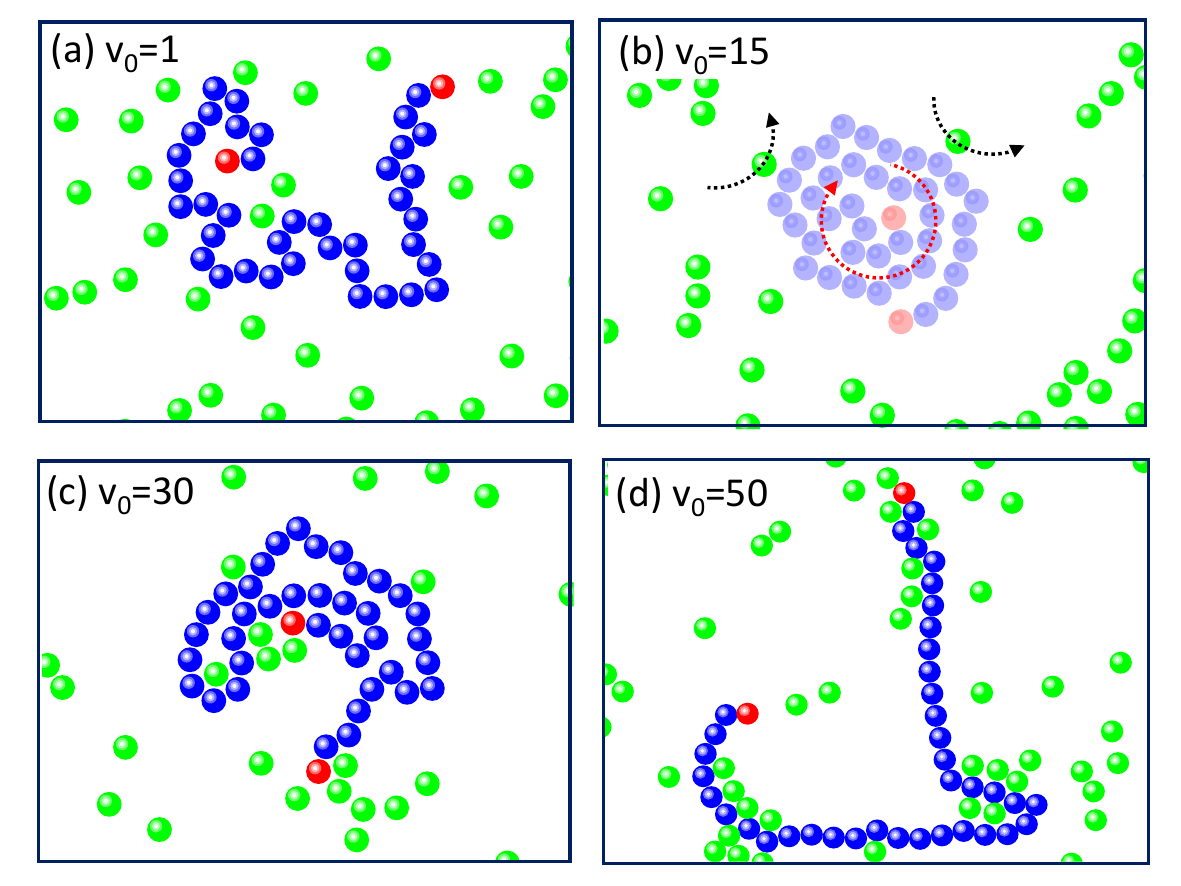}\caption{Typical snapshots for the polymer chain (red and blue) in chiral active
crowders (green) with $\omega=1.5\pi$ and $N=40$ for different active
velocity $v_{0}$. (a) The polymer chain keeps a free state for $v_{0}=1$,
(b) and collapses into a stable compact spiral structure rotating
oppositely to crowders for $v_{0}=15$. (c) The spiral structure is
broken for $v_{0}=30$, (d) and finally the polymer chain expands
for $v_{0}=50$. }
\end{figure}

To quantitatively characterize the phenomena, we have calculated the
radius $R_{g}$ of gyration of the polymer chain defined as $R_{g}=\sqrt{\left\langle R_{g}^{2}\right\rangle }=\sqrt{\frac{1}{2N^{2}}{\displaystyle \sum}_{i,j}(\boldsymbol{r}_{i}-\boldsymbol{r}_{j})^{2}}$.
In Fig.2(a), $R_{g}$ is presented as a function of the velocity $v_{0}$(dash-dotted
line), wherein a clearcut non-monotonic behavior is observed: $R_{g}$
first decreases from $R_{g}^{0}$, the gyration of radius in the absence
of activity ($v_{0}=0$), to a minimum value at $v_{0}\sim15$ corresponding
to the compact spiral cluster shown in Fig.1(b) and then increases
with increasing activity. Finally, $R_{g}$ becomes larger than $R_{g}^{0}$
indicating that the polymer chain swells for large enough activity.
For comparison, the result for achiral crowders, i.e. $\omega=0$,
is also shown (dashed line). Clearly, the behavior is quite distinct
from that for chiral crowders with $\omega=1.5\pi$: $R_{g}$ increases
monotonically with $v_{0}$, in consistent with the results reported
in Ref.\cite{kaiser2014unusual} where it was shown that activity
always leads to swelling of flexible polymer chains. Therefore, chiral
active crowders have very nontrivial effects to the configuration
dynamics of the polymer chain, in particular, they lead to collapse
of the chain to a compact cluster before swelling. 

To get more information about the configuration at different activity
level, we have also measured the probability distribution function
$P\left(R_{g}\right)$, as depicted in Fig.2(b). All the distributions
are unimodal, i.e., the gyration radius peaks around a most probable
value. With increasing activity, the peak first shifts to small values
of $R_{g}$ and then to large values if activity is large, corresponding
to the non-monotonic dependence of $R_{g}$ with $v_{0}$ shown in
Fig.2(a). In addition, one can see that the peak width also shows
non-monotonic variations with increasing $v_{0}$. In particular,
the distribution becomes sharply peaked for $v_{0}\sim15$, wherein
the polymer collapses into a compact cluster. Such small fluctuations
of $R_{g}$ indicates that this spiral cluster is quite stable during
the time evolution. 

\begin{figure}
\centering{}\label{fig.Rg40}\includegraphics[scale=0.7]{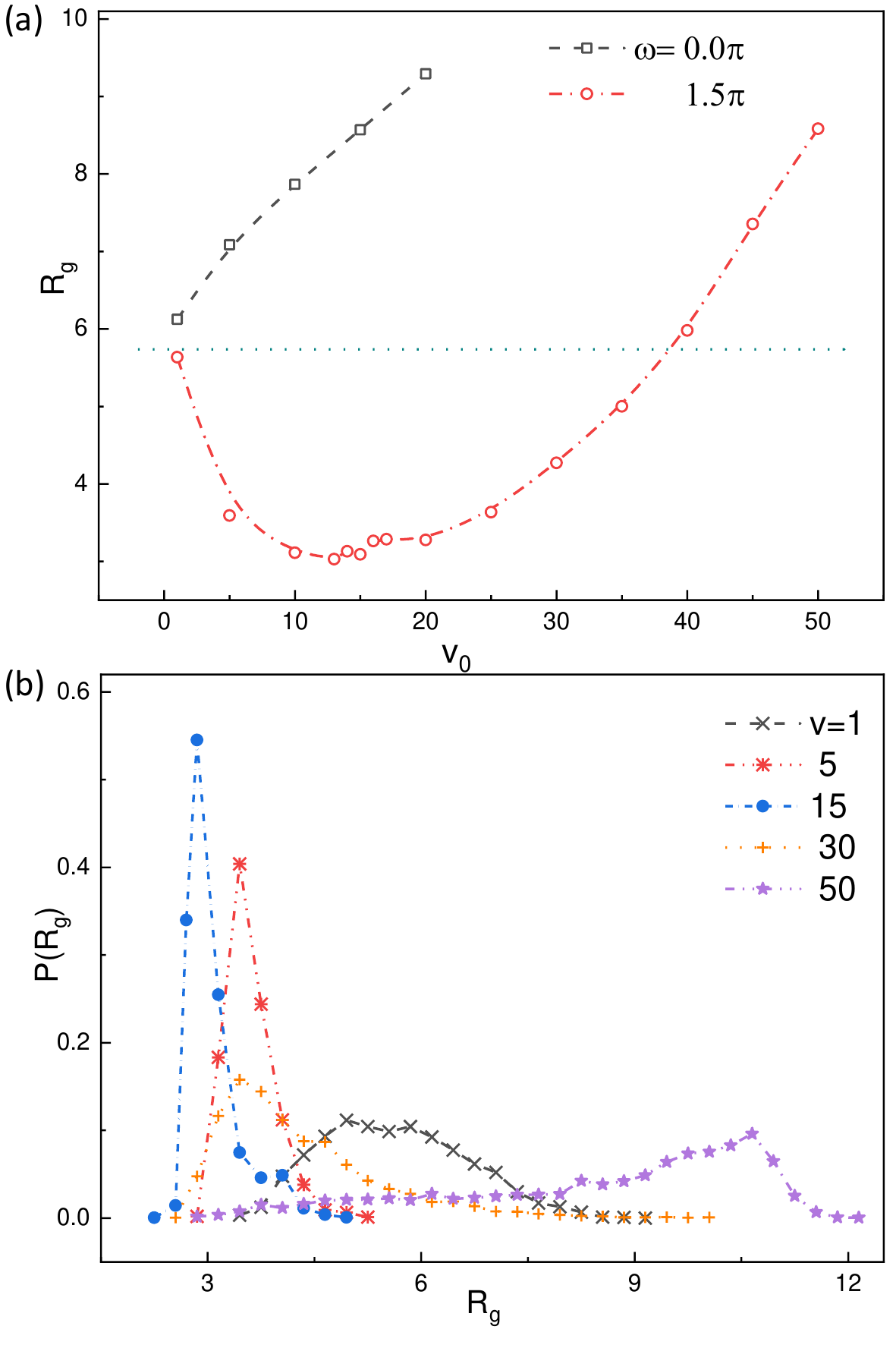}\caption{(a) The radius $R_{g}$ of gyration of the polymer chain with length
$N=40$ as a function of active velocity $v_{0}$ for an achiral active
bath and a chiral one. The horizontal dotted line is $R_{g}^{0}$.
(b) Probability distribution $P(R_{g})$ of the gyration radius $R_{g}$
for the polymer chain with $N=40$ with different active velocity
$v_{0}$and fixed angular velocity $\omega=1.5\pi$.}
\end{figure}

Without doubt, the activity-induced collapse behavior observed above
is due to the chiral feature of the active crowders. As already mentioned
above, the polymer chain would only swell with the increment of crowder
activity $v_{0}$ in the absence of chirality ($\omega=0$). An achiral
active particle, such as the usual active Brownian particle, would
undergo a persistent motion along the original orientation with a
persistence length given by $v_{0}\tau_{p}=v_{0}\left(2D_{r}\right)^{-1}$
before changing its direction randomly according to random rotation.
Therefore, such active particles tend to adhere and aggregate onto
the polymer chain, thus exert a `pulling' force to the polymer beads
resulting in a net expanding effect on the chain length. A particle
will stay aside the polymer chain until its direction changes to an
escape cone and not hindered by peripheral particles. With increasing
activity, the persistence length increases and the force each active
particle exerts onto the chain also increases, leading to more expanded
configuration.

However, such a scenario is not totally true for chiral active particles.
Clearly, an isolated chiral active particle would undergo a deterministic
circular motion with a radius given by $R_{0}=v_{0}/\omega$ if noise
is absent(for $\omega>0$, the particle rotates counter-clockwisely).
Once it collides with the polymer chain, its direction can change
due to this circular motion within a time scale given by $\omega^{-1}$,
unlike the case for achiral particle for which the direction can only
change via random rotation with a characteristic time scale given
by $\tau_{p}$. Such a deterministic circular motion of chiral active
particles would prevent them from adhering to the polymer chain. In
addition, the circular motion of size $R_{0}$ also makes it hard
for the particle to stay in the `interior' part of the chain (locally
concave), which will lead to a difference of pressure between the
`interior' and `exterior' parts of the chain, thus causing the
collapse behavior. 

To get more details of the collapse process, we have investigated
how $R_{g}$ change with time for $v_{0}=15$ (corresponding to the
compact structure shown in Fig.1b) as presented in Fig.3. Correspondingly,
some typical snapshots are also shown. One can see that $R_{g}$ reduces
monotonically (with fluctuations) to the steady state value in a relatively
short period of time. In the early stage, the chain is in an elongated
state (the first snapshot). Shortly, the chain is rolled up from both
ends forming two small connected compact clusters, wherein no crowders
exists in the interior parts. Note that the two ends roll up in the
same clockwise direction (note here the angular velocity of the crowders
$\omega>0,$ i.e., the circular motion is anticlockwise). These two
clusters grow up in size with time until they collide with each other
as shown by the third snapshot. By chance, one cluster will finally
enslave the other and the two clusters combine into one, rotating
clockwisely with one end in and the other end out, as shown by the
last snapshot in Fig.3 which is the same as in Fig.1(b). 

\begin{figure}
\begin{centering}
\includegraphics[scale=0.7]{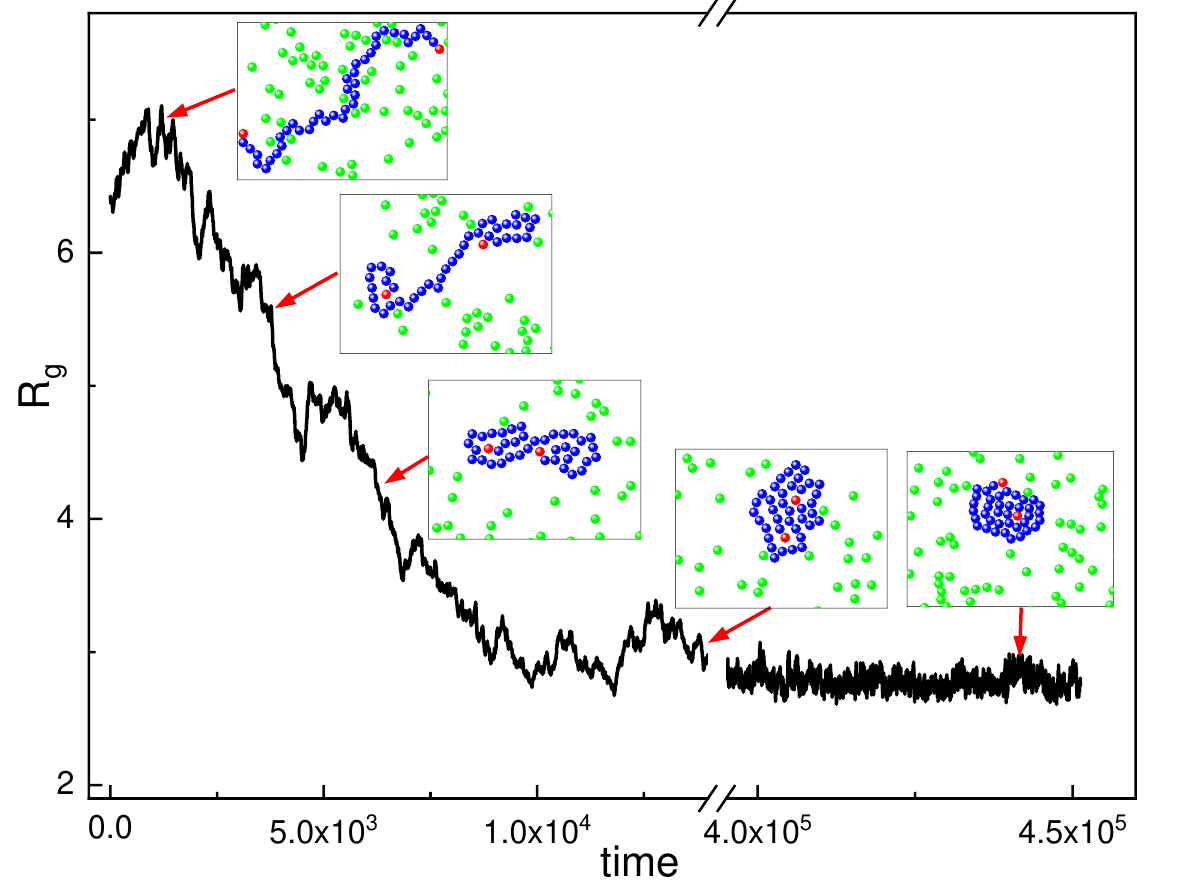}
\par\end{centering}
\caption{Time dependence of the gyration radius $R_{g}$ for a polymer chain
$N=40$ at $v_{0}=15$ and $\omega=1.5\pi$. }
\label{Fig-RgTime}

\end{figure}

The above analysis shows that circular motion is the very reason for
polymer collapse. In addition, the size $R_{0}$ seems to play an
important role. If $R_{0}$ is too small, which is the case for small
$v_{0}$, the effect of chirality is not strong and the crowder particle
acts as a normal achiral one, such that they can still stay in the
interior part of the chain and the pressure difference mentioned in
the last paragraph would be small. In this situation, the collapse
effect would also be small. With the increasing of activity, $R_{0}$
will also increase and fewer particles can stay `inside' the chain,
leading to more strong collapse effect and decreasing $R_{g}$. Of
course, the most collapsed configuration of the chain that could be
reached is the compact one as shown in Fig.1(b).

Note however, increasing particle activity also has another effect,
i.e., the increase of persistence length of the active motion, which
is the reason for polymer stretching as already discussed above. If
the activity is too large, $R_{0}$ would be even larger than the
characteristic length of the chain, and the chain would not `feel'
the circular motion of the particle. Once the particle collide and
adhere to the chain, it will undergo persistence motion along the
chain, just like an achiral active particle. Therefore, the polymer
chain would be elongated again with increasing activity. 

To summarize so far, we understand that dual effects exist for chiral
active crowders on the configuration dynamics of the polymer chain.
On one hand, the circular motion leads to an effect like osmotic pressure
that results in polymer collapse. On the other hand, the persistence
motion would cause stretching of the chain thus leading to swelling.
With increasing activity level $v_{0}$, such two effects will compete
with each other. Firstly, the collapse effect dominates, but finally
the swelling effect will take the dominant role. Consequently, $R_{g}$
shows a non-monotonic dependence on $v_{0}$, namely, it first decreases
to a minimum value and then increases again to a value that is larger
than $R_{g}^{0}$.

\begin{figure}
\begin{centering}
\includegraphics[scale=0.7]{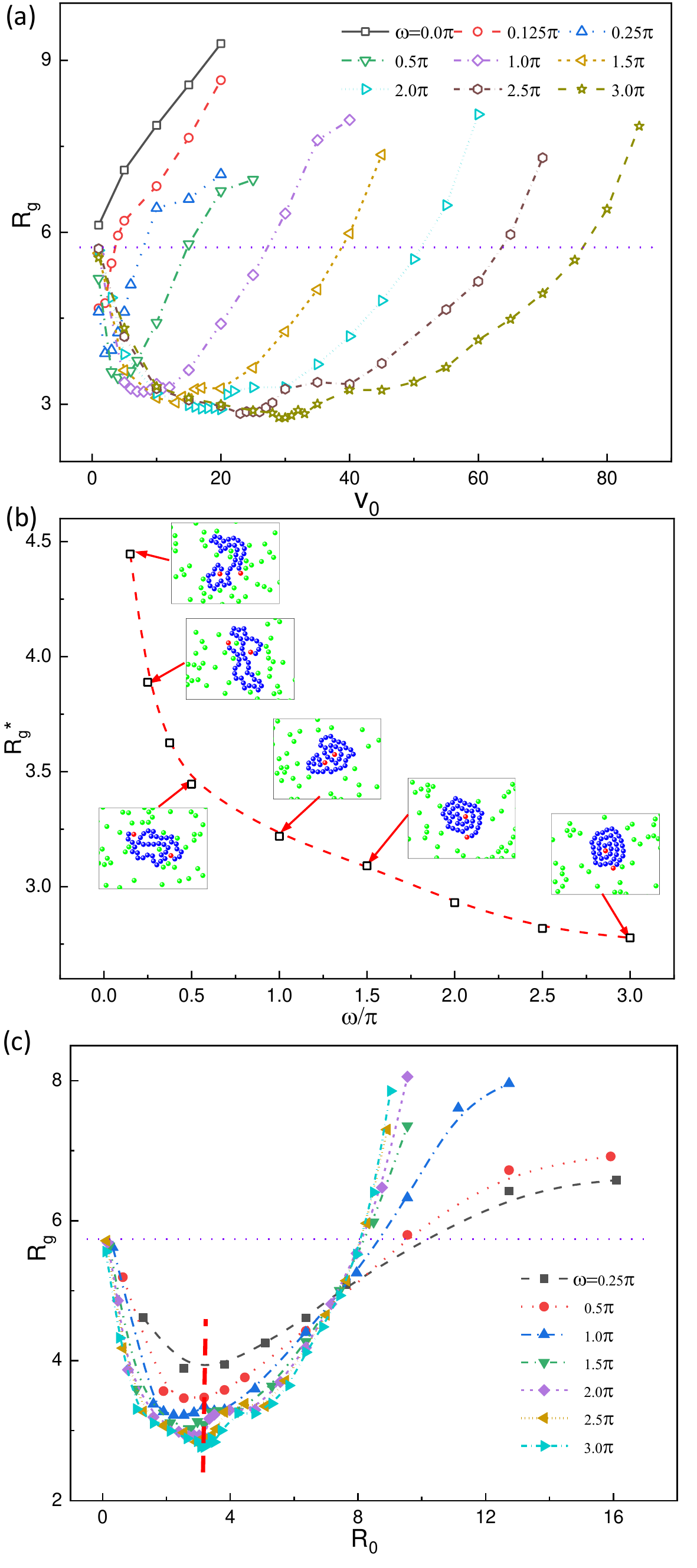}\caption{(a) The radius $R_{g}$ of gyration of the polymer chain with length
$N=40$ as a function of active velocity $v_{0}$ for different angular
velocity $\omega$. (b) Dependence of $R_{g}^{*}$ on angular velocity
$\omega$. (c) The radius $R_{g}$ of gyration of the polymer chain
as a function of the size of the circular motion $R_{0}$ for different
angular velocity $\omega$. The horizontal dotted line in (a) and
(c) is $R_{g}^{0}$.\textcolor{blue}{{} }}
\par\end{centering}
\end{figure}

All above results are obtained for a fixed angular velocity $\omega=1.5\pi$.
One may thus wonder how the findings depend on $\omega$. The results
are presented in Fig.4(a), where the dependences of $R_{g}$ as functions
of $v_{0}$ are shown for a few different values of $\omega$ ranging
from a very small value $\omega=0.125\pi$ to a large one 3.0$\pi$.
One can see that for a very small $\omega$ (say 0.125$\pi$), the
polymer already collapses for a small $v_{0}$($R_{g}<R_{g}^{0}$),
but $R_{g}$ increases with $v_{0}$(the minimum locates at a very
small $v_{0}$ out of the shown range). With increasing $\omega$,
the non-monotonic dependence of $R_{g}$ with $v_{0}$ becomes apparent
and the location of $v_{0}$ for the minimum shifts to larger values.
Meanwhile, the minimum $R_{g}^{*}$ decreases until it saturates to
the smallest value corresponding to a compact cluster, as demonstrated
in Fig.4(b). Also shown in (b) are typical configurations of the chain
with the minimum $R_{g}$. We note here that for small $\omega$,
the most collapsed structure of the chain is not the spiral cluster
as that for large values of $\omega$. If $\omega$ is very small
such that characteristic time scale for deterministic rotation is
larger than the persistence time $\tau_{p}$ of the active motion,
the particle may attach to the chain for a time about $\tau_{p}$
before it rotates out. Consequently, some particles can stay `inside'
the chain as shown in the snapshots in (b) and $R_{g}$ is not that
small. 

The data in Fig.4(a) show that the minimum shifts to larger value
of $v_{0}$ with increasing $\omega$. As already mentioned above,
the size of the circular motion $R_{0}=v_{0}/\omega$ plays an important
role to the collapse behavior of the chain. It suggests us to see
how $R_{g}$ changes with this size parameter $R_{0}$. In Fig.4(c),
$R_{g}$ is drawn as a function of $R_{0}$ instead of $v_{0}$ for
different values of $\omega$. Very interestingly, the locations of
minimum are nearly the same(see the vertical dashed line), i.e., the
chain becomes most collapsed at the same value of $R_{0}$. Therefore,
it is the effective size $R_{0}$ that mainly determines the collapse
behavior of the chain. It should be instructive to perform some theoretical
analysis about how $R_{g}^{*}$ depends on $R_{0}$, but it seems
to be quite hard and beyond the scope of current study. 

In Fig.4(b), $R_{g}^{*}$ reaches a constant value for large values
of $\omega$, corresponding to the close-packed cluster structure
of the chain. Clearly, this constant value of $R_{g}^{*}$ is decided
by the chain length $N$. One may then ask how the above observations
depend on the chain length. At first thought, no qualitative difference
would be expected for different $N$, however, this is not the case.
In Fig.5(a), the dependence of $R_{g}$ on $v_{0}$ is shown for $\omega=1.5\pi$,
with same parameters as in Fig.(2a), but now for $N=60$. Again, one
can see that $R_{g}$ decreases with $v_{0}$ increasing from $v_{0}=1$,
reaches a minimum at $v_{0}\simeq15$ and then increases again, similar
to that observed for $N=40$. But surprisingly, $R_{g}$ shows quite
nontrivial non-monotonic behaviors in the range of moderate values
of $v_{0}$, namely, it undergoes a maximum at $v_{0}\simeq22$ and
another minimum at $v_{0}\simeq30$ before finally increasing monotonically,
which was not observed at all for $N=40$. 

To get more details, we have calculated the distribution of $R_{g}$
at those three extreme values of $v_{0}$ and drawn the typical snapshots
of the chain, as presented in Fig.5(b). At the first minimum $v_{0}\simeq15$,
the distribution is unimodal similar to that for $N=40$, and the
corresponding configuration is also a compact spiral cluster as given
by the inset snapshot. For $v_{0}=22.5$, which is the location for
the maximum of $R_{g}$, the distribution is also unimodal and sharp-peaked
with an average $R_{g}$ larger than that for $v_{0}=15$. Interestingly,
now the typical configuration of the chain is a closed-ring, which
is hollow with some crowder particles moving inside. The ring rotates
in the same direction (anticlockwisely) as the crowders, which is
quite different from the cluster observed at $v_{0}=15$, that rotates
in the opposite direction as the crowders. For $v_{0}=30$ where the
average $R_{g}$ is at a local minimum, however, the distribution
of $R_{g}$ is bimodal with two peaks, one at $R_{g}\simeq3.5$ that
is close to the peak position for $v_{0}=15$, and the other at $R_{g}\simeq5.5$
close to the peak position for $v_{0}=22.5$. Correspondingly, typical
snapshots for the two peaks are shown in the same figure, wherein
the first peak is related to the cluster structure and the second
one to the ring-structure. Watching the dynamic process, we find that
the polymer chain actually oscillates between these two structures.
In Fig.5(c), we have drawn the time dependence of $R_{g}$ for $v_{0}=30$,
where one can observe quite obvious oscillation behavior. We note
here that similar oscillation behaviors are observed for a quite range
of $v_{0}$ to the right side of the maximum. 

\begin{figure}
\centering{}\includegraphics[scale=0.7]{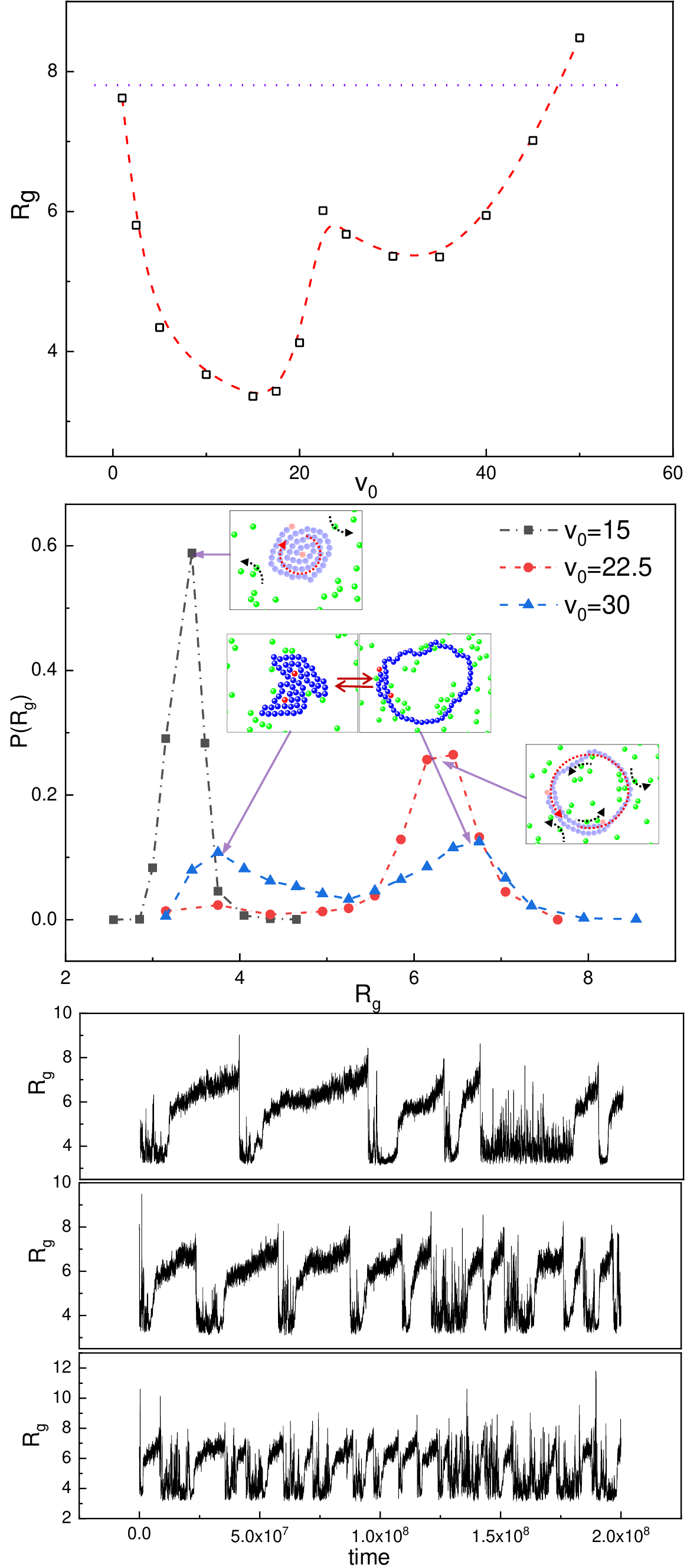}\caption{(a) The radius $R_{g}$ of gyration of the polymer chain with length
$N=60$ as a function of active velocity $v_{0}$. The horizontal
dotted line is $R_{g}^{0}$. (b) Probability distribution $P(R_{g})$
of the gyration radius $R_{g}$ for various active velocity $v_{0}$.
(c) Time dependence of the gyration radius $R_{g}$ for a polymer
chain $N=60$ for several active velocity $v_{0}$. The angular velocity
is fixed as $\omega=1.5\pi$.}
\end{figure}

Here we would like to point out that the rotation directions of the
cluster (for $v_{0}=15$) and the ring (for $v_{0}=22.5$) observed
here are consistent with those observations in Ref.\cite{liao2018clustering}
and Ref. \cite{chen2017rotational}. In Ref.\cite{liao2018clustering},
the authors studied\textcolor{blue}{{} }\textcolor{black}{the phase
separation of chiral active crowders}, finding that the passive cluster
rotates in the opposite direction to that of the\textcolor{blue}{{}
}\textcolor{black}{chiral }active particles. There the authors argued
that the passive cluster with non-smooth boundary formed a `gear',
which was the reason for opposite rotating behavior. The same reasoning
also works for the spiral cluster formed by the polymer chain here.
While in Ref.\cite{chen2017rotational}, Chen et al. studied the behavior
of a closed ring (note the ends are connected there) with some chiral
active particles inside. They found that the ring rotated continuously
in the same direction as the active particles for a proper level of
activity, which is similar to what we observed for $v_{0}=22.5$ here.
Note that the ring-configuration observed in our present study is
dynamically formed during the process, with active particles dynamically
going in and out in a stationary way. 

The results observed for $N=60$ further demonstrate that nontrivial
roles played by the chiral crowders on the chain dynamics. The formation
of the compact cluster structure at the first minimum can be understood
in the similar way as that for $N=40$, which is due to the circular
motion of the chiral particle. The ring configuration observed at
a moderate value of $v_{0}=22.5$, which was not observed for $N=40$,
is also closely related to the circular motion of the chiral crowders.
For this ring to be formed, the chain length must be large enough
to support the circular motion of crowders inside, thus it can not
be observed if the chain is too short. The crowders moving inside
the ring collide with the polymer beads and expand the ring, making
the ring rotates in the same direction as the circular motion as mentioned
above. The crowders outside the ring, however, tend to collapse the
ring and rotate the ring in the opposite direction. If the activity
level is small, the circular motion size $R_{0}$ is not large enough
to sustain the ring from inside. If activity is too large, on the
other hand, the pressure outside the ring would dominate (since the
particles inside the ring is limited) causing instability of the ring.
The competition between the different effects inside/outside the ring
thus leads to the non-monotonic as well as oscillating behaviors observed
for the moderate levels of $v_{0}$. If $v_{0}$ is very large, the
stretching effect would dominate finally resulting in monotonic increasing
of $R_{g}$ again. We also note here that these behaviors observed
for $N=60$ are also present for longer chains except that the locations
of $v_{0}$ for the minimum or maximum are different. Therefore, the
results reported here for $N=40$ and 60 already outline the main
dynamics of a flexible chain in chiral active crowders. 

Finally, we have also investigated the results for $N=60$ for different
values of $\omega$. As in Fig.4(c), we draw $R_{g}$ as a function
of $R_{0}=v_{0}/\omega$, as shown in Fig.6. A few remarks can be
made regarding the results. Firstly, the maximum of $R_{g}$ is not
apparently observed for small $\omega$(see for instance $\omega=0.5\pi$
and $1.0\pi$), indicating that the ring structure can not be stably
formed. Secondly, we also find that the positions of $R_{0}$ for
the first minimum (for cluster structure) are still nearly the same
similar to the case for $N=40$. More interestingly, once the maximum
(for ring structure) exists, it also takes place at nearly the same
value of $R_{0}$. Finally, the oscillation behavior is robust if
$\omega$ is not small, and the location of the second minimum also
takes place at nearly same $R_{0}$. All these findings demonstrate
the key role of the circular motion, which is the very characteristics
of chiral active particles, and underline a type of size effect which
deserves more study in future works. 

\begin{figure}
\centering{}\label{fig.Rg60}\includegraphics[scale=0.7]{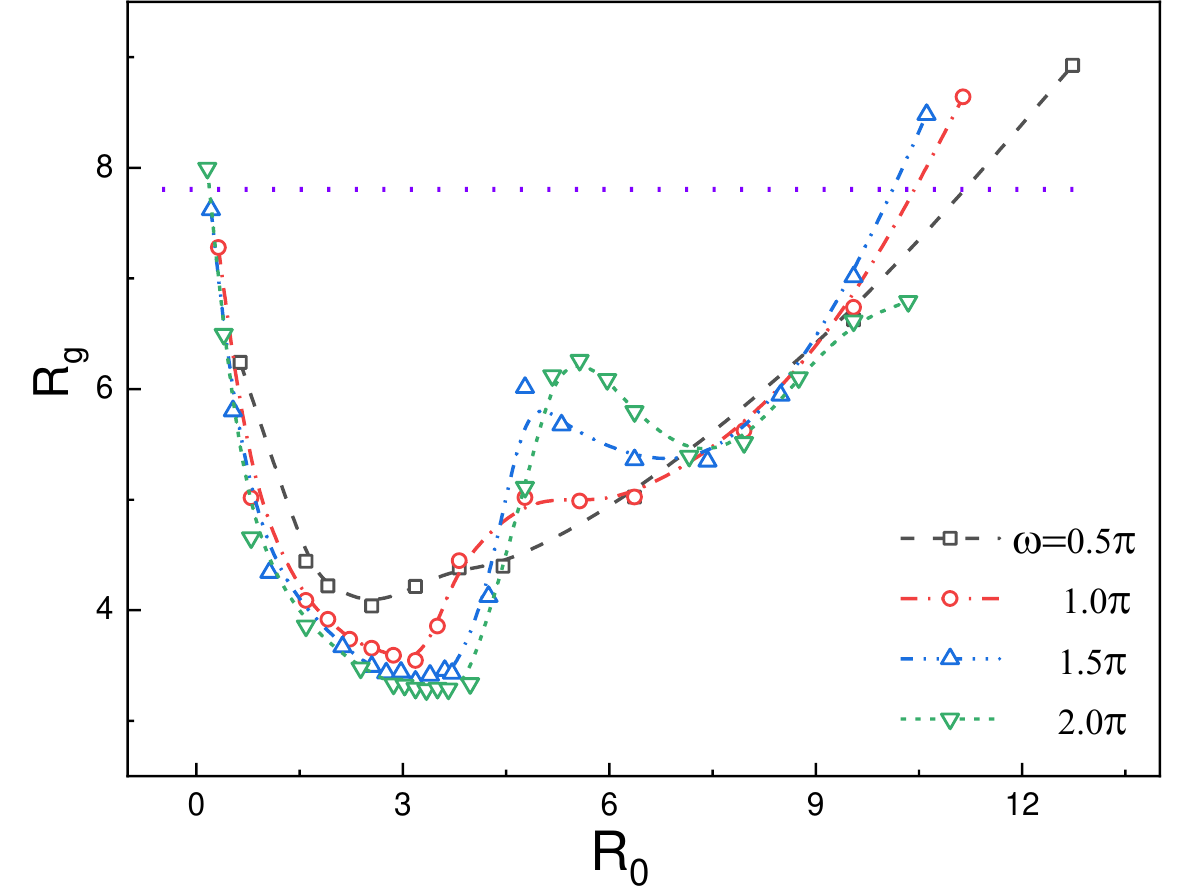}\caption{The radius $R_{g}$ of gyration of the polymer chain as a function
of the size of the circular motion $R_{0}$ for different angular
velocity $\omega$. }
\end{figure}

\section{CONCLUSION}

In conclusion, we have used Langevin dynamics simulation to investigate
the \textcolor{black}{configuration dynamics of }a flexible linear
polymer chain in a bath of active crowder particles with dynamic chirality,
i.e., each particle undergoes self-propelled active motion with velocity
$v_{0}$ along an orientation which changes via random rotational
diffusion besides a deterministic circular motion with angular velocity
$\omega$. Different from the results for chain dynamics in a purely
achiral active bath where the chain swells monotonically with increasing
$v_{0}$, it was found that the average radius of gyration $R_{g}$
of the chain shows quite interesting non-monotonic dependences on
$v_{0}$ if $\omega$ is observably nonzero. In particular, the chain
collapses for small values of $v_{0}$ while it swells if $v_{0}$
is large enough, such that $R_{g}$ reaches a minimum value $R_{g}^{*}$
at a moderate $v_{0}^{*}$. Interestingly, the chain is suppressed
to a rotating compact cluster at this minimum if $\omega$ is not
small, of which the rotation direction is opposite to the crowder
particles. In addition, we found that the size $R_{0}=v_{0}/\omega$
of the circular motion plays a subtle role, in that the minima for
different values of $\omega$ take places at nearly the same value
of $R_{0}$. Our analysis revealed that chiral active bath can result
in two-fold effects on polymer\textcolor{black}{{} configuration dynamics:
one of which is that }the persistence motion of the active particles
leads to stretching of the polymer chain and makes the chain swell,
and the other is that\textcolor{black}{{} }the circular motion of the
particle leads to an osmotic-pressure-like effect that results in
polymer collapse. It is the competition between these two effects
that results in the nontrivial dependence of chain dynamics on particle
activity. If the chain is long enough, even more interesting phenomena
can be observed if $v_{0}$ is larger than $v_{0}^{*}$: the chain
can form a hollow closed-ring rotating with the same direction as
that of the crowders, also taking place at nearly the same values
of $R_{0}$, and the ring may lose stability if $v_{0}$ is even larger,
leading to interesting oscillating behavior between the closed-ring
and the compact cluster. As a consequence, for relatively long chains,
$R_{g}$ undergoes two minima as $v_{0}$ increases, with a maximum
between them. Our results clearly demonstrate that chain dynamics
in a chiral active bath is considerably different from that in a normal
achiral one and highlight the nontrivial roles played by the particle
chirality. Our work may shed new lights on the study of active particles
systems since chiral active baths are of ubiquitous importance in
real biological active systems.
\begin{acknowledgments}
This work is supported by MOST(2016YFA0400904, 2018YFA0208702), by
NSFC (21973085, 21833007, 21790350, 21673212, 21521001, 21473165),
by the Fundamental Research Funds for the Central Universities (WK2340000074),
and Anhui Initiative in Quantum Information Technologies (AHY090200).
\end{acknowledgments}

\end{document}